\documentclass[11pt]{article}
\input epsf.tex
\newdimen\SaveWidth \SaveWidth=\textwidth
\newdimen\SaveHeight \SaveHeight=\textheight
\advance\SaveWidth by -\textwidth
\advance\SaveHeight by -\textheight

\divide\SaveWidth by 2
\divide\SaveHeight by 2
\advance\hoffset by \SaveWidth
\advance\voffset by \SaveHeight

\def\ie{\it i.e.}

\def\nc{\rm noncommutative}
\def\ncg{\rm noncommutative~geometry}

\def\cpviolng{CP ~\rm {violating}}

\def\etal{{\it et~al.}}

\def\abs#1{\left| #1\right|}

\let\badcite=\cite
\def\cite{~\badcite}

\def\slashchar#1{\setbox0=\hbox{$#1$}           
   \dimen0=\wd0                                 
   \setbox1=\hbox{/} \dimen1=\wd1               
   \ifdim\dimen0>\dimen1                        
      \rlap{\hbox to \dimen0{\hfil/\hfil}}      
      #1                                        
   \else                                        
      \rlap{\hbox to \dimen1{\hfil$#1$\hfil}}   
      /                                         
   \fi} 
    \def\slashword#1{\setbox0=\hbox{$#1$}        
  \dimen0=\wd0                                   
   \setbox1=\hbox{/} \dimen1=\wd1                
   \ifdim\dimen0>\dimen1                         
      \rlap{\hbox to \dimen0{\hfil\bf---\hfil}} %
      #1                                         %
   \else                                         
      \rlap{\hbox to \dimen1{\hfil$#1$\hfil}}    
      /                                          
    \fi}                                         %

\catcode`@=11
\newdimen\vbigd@men                             

\def\vbig#1#2{{\vbigd@men=#2\divide\vbigd@men by 2%
   \hbox{$\left#1\vbox to \vbigd@men{}\right.\n@space$}}}
\catcode`@=12

\catcode`@=11
\def\citenum#1{\csname b@#1\endcsname}
\catcode`@=12

\def\dofig#1#2{\centerline{\epsfxsize=#1\epsfbox{#2}}}
\begin{document}
\begin{titlepage}
\rightline{LBNL-47750}

\bigskip\bigskip

\begin{center}{\Large\bf\boldmath
$(g-2)_\mu$ from Noncommutative Geometry\footnotemark \\}
\end{center}
\footnotetext{ This work was supported by the Director, 
Office of Science, Office
of Basic Energy Services, of the U.S. Department of Energy under
Contract DE-AC03-76SF0098.
}
\bigskip
\centerline{\bf N. Kersting}
\centerline{{\it Lawrence Berkeley National Laboratory, Berkeley, CA}}
\bigskip

\begin{abstract}

	This brief Letter demonstrates that effects from a  $\nc$ space-time geometry
	will measurably affect the value of $(g-2)_\mu$ inferred from the decay of
	the muon to an electron plus two neutrinos. If the scale of
	noncommutivity is ${\cal O}(TeV)$, the alteration of the $V-A$ structure
	of the lepton-lepton-W vertex is sufficient to shift the inferred value
	of  $(g-2)_\mu$  to one part in $10^8$. This may account for the
	recently reported $2.6 \sigma$ discrepancy between the BNL measurement 
	 $a_{expt} = 11659202(14)(6) \times 10^{-10}$ and the Standard Model prediction 
	 $a_{SM} = 11659159.6(6.7) \times 10^{-10}$. 

\bigskip        

\end{abstract}

\newpage
\pagestyle{empty}

\end{titlepage}

\section*{Introduction}
\label{sec:intro}

The measurement of the anomalous magnetic moment of the muon, $a_\mu \equiv (g-2)_\mu$,
has undergone continual refinement (for history
and experimental details, see\cite{muonedm},\cite{muonold})
to the point where   $a_{\mu}$ is now very precisely known\cite{bnl}:
\begin{equation}
a_{\mu}^{expt} = 11659202(14) \cdot 10^{-10}
\end{equation}
The experimental technique employs muons trapped in a storage ring. A 
uniform magnetic field $B$ is applied perpendicular to the orbit of the 
muons; hence the muon spin will precess. The signal is a discrepancy between
the observed precession and cyclotron frequencies.
Precession of the muon spin is determined indirectly from the decay
$\mu \to e~{\overline \nu}_e ~ \nu_\mu$. Electrons emerge from
the decay vertex with a characteristic angular distribution which
in the Standard Model (SM) has the following form in the rest frame of the muon:
\begin{equation}
\label{asymm}
dP(y,\phi) = n(y) (1 + A(y) cos(\phi))dy d(cos(\phi))
\end{equation}
where $\phi$ is the angle between the momentum of the electron
e and the spin of the muon, $y = 2 p_e/m_\mu$ measures the fraction of the maximum 
available energy which the electron carries, and $n(y),A(y)$ are 
particular functions which peak at $y=1$. The detectors (positioned 
along the perimeter of the ring) accept the passage of only the highest
energy electrons in order to maximize the angular asymmetry 
in (\ref{asymm}). In this way, the electron count rate is
modulated at the frequency $a_\mu e B/(2\pi m c)$. 

The leading theoretical prediction of $a_\mu$ in the SM is
$a_{\mu}^{SM} =  11659159.6(6.7) \cdot 10^{-10}$ \cite{theory}
which leads to a $2.6~\sigma$
deviation from the data:
\begin{equation}
\label{deviate}
a_\mu^{expt} - a_\mu^{SM} =  43(16) \cdot 10^{-10}
\end{equation}
If this discrepancy persists as more data arrives and theoretical uncertainties improve,
then there is a clear signal of new physics. Many proposals to account for this
discrepancy have already appeared
in the literature.\footnotemark 
\footnotetext{for a partial list, see \cite{muonpapers}}

This letter is a consideration of a novel effect on the measurement of
$a_\mu$ from \newline $\ncg$, a theory in which the coordinates of spacetime
become noncommuting operators: $[\widehat{x}_\mu,\widehat{x}_\nu]=i \theta_{\mu \nu}$.
There is an extensive collection of papers devoted to both the theoretical foundations
of $\ncg$\cite{found1,found2,found3, found4,found5,found6} and 
its phenomenology\cite{phen1,phen2,phen3,phen4}. 
The reader may  consult the above references for a more thorough understanding of
the $\nc$ quantum field theory underlying the present calculation. We will 
employ perturbation theory in leading powers of the dimensionful matrix of 
parameters $\theta_{\mu \nu}$ in accord with the work done in \cite{phen4}.

\section*{Preliminaries}
\label{sec:calc}

Although $a_\mu$ does receive a sizable contribution from $\ncg$, 
it is a {\it constant} contribution\cite{phen3}, $\ie$ the
interaction with the external magnetic field $\Delta E \sim B_i \theta_{jk} \epsilon^{ijk}$
is independent of the muon spin, and therefore the
experiment described above is not sensitive to this perturbation of $a_\mu$. 

\begin{figure}[t]
\dofig{4.00in}{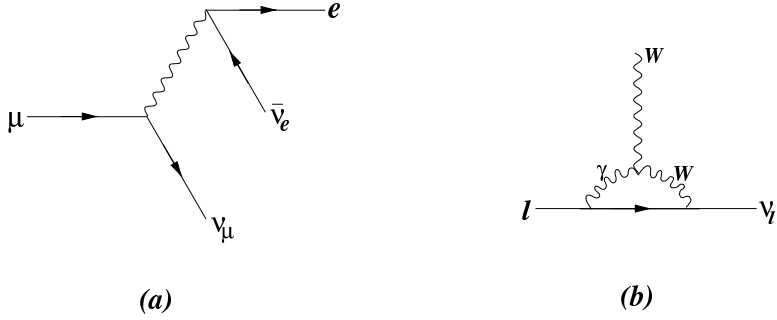}
\caption{(a) The muon decay to an electron plus two neutrinos.
Each vertex receives a $\nc$ loop correction (b) (with $l=e,\mu$) 
which upsets the electron's angular distribution. 
 \label{muonfig} }
\end{figure}

The effect of $\ncg$ on this measurement does however enter in the manner in
which the muon spin is measured in its decay.
Each of the W-boson vertices in the decay diagram Fig.\ref{muonfig}(a)  
receives corrections from $\ncg$ at the one loop level, as shown 
in Fig.\ref{muonfig}(b). One might expect such corrections to be
negligable, but in fact the loop integral in  Fig.\ref{muonfig}(b) 
involves $\theta$-dependent vertices which lead to integrals of
the form
\begin{equation}
\label{loop}
\int {\frac{d^4 k}{16 \pi^2} \frac{e^{i p \cdot \theta \cdot q}}{k^4}}
\end{equation}
for loop momenta much larger than the external momenta $p,q$. 
In the limit $\abs{\theta}\to 0$ the integral (\ref{loop}) formally 
diverges so one has to renormalize carefully (see \cite{phen4} 
for a discussion of this point). The generic size of the 
$\nc$ contribution will be 
$\frac{\alpha}{16 \pi^2}\abs{p_{\mu}^2 \theta}ln\abs{p_{\mu}^2 \theta}$
which for fast muons ($p_{\mu} \approx 3~GeV$ at BNL) and low scales of
noncommutivity ($\abs{\theta} \approx (1~TeV)^{-2}$) gives a suppression
factor of ${\cal O}(10^{-8})$ relative  to the tree level decay diagram. 
Since the current deviation of the SM prediction from experiment in 
(\ref{deviate}) is of this size, we see that $\nc$ effects cannot
be neglected on the basis of their magnitude. 

More importantly, the appearance of the antisymmetric object $\theta_{\mu \nu}$
in the decay amplitude leads to combinations of the muon and electron spins 
and momenta which alter the modulation frequency of the decay rate (\ref{asymm}).
Specifically, one anticipates factors of 
($\overrightarrow{p_e} \cdot \overrightarrow{s}_\mu)
(\overrightarrow{p_e} \cdot \theta \cdot \overrightarrow{s}_\mu)$
which for electron momenta close to their kinetmatical limit ($\ie$ $y=1$)
behaves like $cos(\phi)sin(\phi)$. In what follows we explicitly demonstrate these
terms exist in the decay rate.

\section*{The Calculation\footnotemark}
\footnotetext{for an excellent treatment of the corresponding SM 
	calculation, see \cite{muondecay}}
\label{sec:calc}

\begin{figure}[t]
\dofig{2.00in}{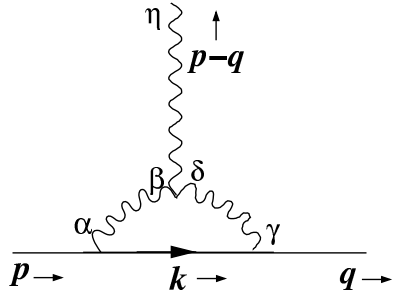}
\caption{Variables defined in the loop calculation}
 \label{loopfig} 
\end{figure}

Define the muon decay amplitude 
\begin{equation}
{\cal M} = \frac{G_F}{\sqrt{2}} \overline{u}_e (C_i {\cal O}_i^\alpha)v_1
	\overline{u}_2 (C_j' {\cal O}_j'^\alpha)u_\mu
\end{equation}
involving the electron, muon, and neutrino($1,2$) spinors and the
most general set of operators at the interaction vertices, $ {\cal O}_i$
($i \subset \{S,P,A,V,T \}$) which may depend on momenta. The muon decay
rate is proportional to the squared matrix element
\begin{equation}
\begin{array}{l}
\abs{\cal M}^2 =  \frac{G_F^2}{2} T_e T_\mu \\
T_e \equiv tr\left(\overline{u}_e (C_i {\cal O}_i^\alpha)v_1
         \overline{v_1}(C^*_j {\cal O}_j^\beta)u_e \right) \\
T_\mu \equiv  tr\left(\overline{u}_2 (C_k {\cal O}_{k,\alpha})u_\mu
         \overline{u_\mu}(C^*_l {\cal O}_{l,\beta})u_2 \right) \\
\end{array}
\end{equation}
This is a product of two terms: the electron trace $T_e$ and the muon 
trace $T_\mu$.
If $\theta$ were zero, all operators would be of the standard $V-A$ form, and
the traces would be 
\begin{equation}
\label{traces}
\begin{array}{ll}
T_e (SM) = & 4 \left( q_1^\alpha p_e^\beta +q_1^\beta p_e^\alpha
	- (q_1 \cdot p_e) g^{\alpha \beta} + 
	iq_1^\gamma p_e^\delta \epsilon^{\alpha\beta\gamma\delta} \right) \\
T_\mu (SM) = & 4 \left( q_2^\alpha p_\mu^\beta +q_2^\beta p_\mu^\alpha
	- (q_2 \cdot p_\mu) g^{\alpha \beta} + 
	iq_2^\gamma p_\mu^\delta \epsilon^{\alpha\beta\gamma\delta} \right) \\
& - 4m  \left( q_2^\alpha s_\mu^\beta +q_2^\beta s_\mu^\alpha
	- (q_2 \cdot s_\mu) g^{\alpha \beta} + 
	iq_2^\gamma s_\mu^\delta \epsilon^{\alpha\beta\gamma\delta} \right)\\
\end{array}
\end{equation}
where $m$ is the muon mass and we neglect the mass of the electron in this 
and all that follows. 
The lowest 
order contribution from $\ncg$ will be proportional to one power of
$\theta$, so to extract it one calculates the contribution to $\abs{\cal M}^2$
from each way it is possible to change one $V-A$ operator into a $\nc$ one,
giving altogether twenty $ {\cal O}(\theta)$ terms in  $\abs{\cal M}^2$. 
To find the precise form of these operators, we next calculate the loop.
In Fig.\ref{loopfig} we show the loop with incoming charged lepton momentum $p$
and outgoing neutrino momentum $q$. The loop amplitude is
\begin{equation}
\begin{array}{ll}
{\cal M}_{loop} = & \int \frac{d^4 k}{(2 \pi)^4} 
\overline{u}_e(q)[-ig\gamma^\gamma(1-\gamma_5)]
\frac{-i g_{\gamma \delta}}{(k-q)^2 - m_W^2}
\frac{i}{\slashchar{k} - m}[-ie \gamma^\alpha]\frac{-i g_{\alpha \beta}}{(p-k)^2} u(p) \\
& \times g [g^{\eta \beta}(q+k-2p)^\delta + g^{\beta \delta}(q+p-2k)^\eta
   + g^{\eta \delta}(k+p-2q)^\beta] exp[i k \cdot \theta \cdot (p-q)] \\
\end{array}
\end{equation}
which becomes
\begin{equation}
\begin{array}{l}
\overline{u}_e(q) g^2 e\int \left( \frac{d^4 k}{(2 \pi)^4}
	\frac{N_1^\eta + N_2^\eta + N_3^\eta}
{(k^2-m^2)(p-k)^2((k-q)^2 - m_W^2)}e^{i k \cdot \theta \cdot (p-q)} \right)u(p) \\
N_1^\eta = (\slashchar{q}+\slashchar{k}-2\slashchar{p})(1-\gamma_5)
		( \slashchar{k}+m)\gamma^\eta \\
N_2^\eta = \gamma^\beta(1-\gamma_5)( \slashchar{k}+m) \gamma^\beta
	   (q+k-2p)^\eta \\
N_3^\eta =  \gamma^\eta(1-\gamma_5)(\slashchar{p}+\slashchar{k}-2\slashchar{q}) \\
\end{array}
\end{equation}
Now using the on-shell condition $\overline{u}_e(p)\slashchar{p} = m\overline{u}_e(p)$
and only retaining terms which couple $\theta_{\mu \nu}$ to the overall Dirac
structure \footnotemark 
\footnotetext{$\ie$ terms containing $k^\eta$ or $\slashchar{k}$, since 
 $\theta$ needs to be contracted with the electron or muon spin}
we arrive at
\begin{equation}
\begin{array}{l}
N_1^\eta \to 2m\slashchar{k}(1+\gamma_5)\gamma^\eta 
	- 2\slashchar{k}p^\eta(1+\gamma_5)\\
N_2^\eta \to m k^\eta(1+\gamma_5) -2 k^\eta \slashchar{k} (1-\gamma_5) \\
N_3^\eta  \to m  \gamma^\eta(1-\gamma_5)\slashchar{k} \\
\end{array}
\end{equation}
Of the above terms in the numerator, the dominant one is the tensor piece
of $N_2^\eta$, $\ie$  the one proportional to $k^\eta \slashchar{k}$, since
it has the most powers of $k$. To compute its effect, we consider first
the alteration of the electron trace, keeping the $V-A$ vertices of the
muon trace intact. This tensor part of the electron trace $T_e$ is
\begin{equation}
\begin{array}{l}
T_e = tr\left(\slashchar{p_e}(1-\gamma_5\slashchar{s_e})\gamma^\mu \gamma^\alpha
	\theta_{\mu \rho} (p_e - q_1)^\rho \slashchar{q_1}
	 \gamma^\beta(1-\gamma_5)\right) \\
	+ tr \left(\slashchar{p_e}(1-\gamma_5\slashchar{s_e})
	\gamma^\alpha(1-\gamma_5) \slashchar{q_1}
	\gamma^\mu \gamma^\beta\theta_{\mu \rho} (p_e - q_1)^\rho \right) \\
	\times \frac{g^2 e}{16 \pi^2}ln\abs{m_\mu^2 \theta}
\end{array}
\end{equation}
which, after some Dirac algebra, dotting into the SM muon trace (\ref{traces}),
  and integration over the neutrino momenta $q_{1,2}$
(since these are not observed) gives
\begin{equation}
\abs{\cal M}^2 \supset  \frac{G_F^2 g^2 e m_\mu^6}{64 \pi}
			ln\abs{m_\mu^2 \theta}(s_e \cdot \hat{p}_e)
			(s_\mu \cdot \theta \hat{p}_e)
\end{equation}
The other half of the calculation, keeping the electron trace fixed
and inserting $\theta$-dependent operators into the muon trace,
yields a very similar result. 
For high electron momenta, the muon neutrino and electron antineutrino 
momenta are approximately opposite that of the  electron, forcing the spin of the
electron to match the spin of the muon. In this case the product
$(s_e \cdot \hat{p}_e)(s_\mu \cdot \theta \cdot \hat{p}_e)$ becomes
approximately $cos(\phi)sin(\phi)$  since $\overrightarrow{s}_e 
\approx \overrightarrow{s}_\mu$ and  $\theta_{\mu \nu}$ is
antisymmetric. This upsets the $cos(\phi)$ angular dependence that
the SM predicts in (\ref{asymm}) 
potentially at the level of 1 part in $10^8$.

\section{Concluding Remarks}

It is interesting not only that $\ncg$ can account for the 
recent measurement of $a_\mu$ if the scale of noncommutivity
is of the order of $1~TeV$, but also that a $\nc$ spacetime
at this energy can account for $\epsilon_K$ and possibly some
of the $\cpviolng$ observables in $B$-meson physics \cite{phen4}.
The caveat however is that $\theta_{\mu \nu}$, being an intrinsically
directional object, is subject to being averaged away if
experiments collect and average data over time scales of days or
longer due to the rotation of the Earth. In a storage ring such as
the one at BNL, the circulation of the
muons at their cyclotron frequency introduces an additional averaging of the
components of $\theta$, so some of the effects of $\ncg$ are
bound to be projected away.
Nonetheless, it is hoped that experimenters will look for a time-varying
effect in the data for $a_\mu$ which would be a definite positive
signal of $\ncg$.

\section*{Acknowledgements}
 This work was supported by the Director, Office of Science, Office
of Basic Energy Services, of the U.S. Department of Energy under
Contract DE-AC03-76SF0098.

\pagebreak

\end{document}